\documentclass[conference]{IEEEtran}
\IEEEoverridecommandlockouts
\usepackage{cite}
\usepackage{amsmath,amssymb,amsfonts}
\usepackage{graphicx}
\usepackage{textcomp}
\usepackage{xcolor}
\usepackage{balance}
\usepackage{cite}
\usepackage{hyperref}
\usepackage{amsmath,amssymb,amsfonts}
\usepackage{amsmath}
\usepackage{amsthm}

\usepackage{graphicx}
\usepackage{textcomp}
\usepackage{xcolor}
\usepackage{graphicx}
\usepackage{float}
\usepackage{subfigure}
\usepackage{amsmath}
\usepackage{amsfonts,amssymb}
\usepackage{mathrsfs}
\usepackage{mathtools}
\usepackage{bm}
\usepackage{multirow}
\usepackage{array}
\usepackage{amssymb}
\usepackage{amsmath}
\usepackage{cite}
\usepackage{url}
\usepackage{xcolor}
\usepackage{cite,graphicx,amsmath,amssymb}
\usepackage{subfigure}
\usepackage{fancyhdr}
\usepackage{mdwmath}
\usepackage{mdwtab}
\usepackage{caption}
\usepackage{amsthm}
\usepackage{setspace}
\usepackage{bm}
\usepackage{mathtools}
\usepackage{dsfont}
\usepackage{bbm}
\usepackage{algorithm}
\usepackage{algorithmic}
\newtheorem{remark}{Remark}
\newtheorem{theorem}{Theorem}

\newtheorem{lemma}{Lemma}

\newtheorem{corollary}{Corollary}

\makeatletter
\newcommand{\biggg}{\bBigg@{3}}
\newcommand{\Biggg}{\bBigg@{3.5}}
\makeatother
\def\BibTeX{{\rm B\kern-.05em{\sc i\kern-.025em b}\kern-.08em
    T\kern-.1667em\lower.7ex\hbox{E}\kern-.125emX}}
\begin{document}

\title{Capacity Scaling Law in Massive MIMO with Antenna Selection}

\author{\IEEEauthorblockN{Chongjun Ouyang, Hao Xu, Xujie Zang, and Hongwen Yang}
School of Information and Communication Engineering\\
Beijing University of Posts and Telecommunications, Beijing, 100876, China\\
\{DragonAim, Xu\_Hao, zangxj, yanghong\}@bupt.edu.cn}

\maketitle

\begin{abstract}
Antenna selection is capable of handling the cost and complexity issues in massive multiple-input multiple-output (MIMO) channels. The sum-rate capacity of a multiuser massive MIMO uplink channel is characterized under the Nakagami fading. A mathematically tractable sum-rate capacity upper bound is derived for the considered system. Moreover, for a sufficiently large base station (BS) antenna number, a deterministic equivalent (DE) of the sum-rate bound is derived. Based on this DE, the sum-rate capacity is shown to grow double logarithmically with the number of BS antennas. The validity of the analytical result is confirmed by numerical experiments.
\end{abstract}


\section{Introduction}
To meet the immense demand for mobile data traffic, massive multiple-input multiple-output (MIMO) has emerged as a key enabling technology in wireless networks due to its promising spectral efficiency performance \cite{Marzetta2016_CM}. By deploying a large-scale antenna array at the base station (BS), massive MIMO offers unprecedented degrees of freedom (DoFs) to the wireless channel. Yet, to reap the maximal transmission rate, each BS antenna element needs to be accompanied with a dedicated radio frequency (RF) chain, which significantly increases the hardware cost, power consumption, and signal processing complexity of the system \cite{Lu2014_JSTSP}. To alleviate these issues, numerous methods have been recently proposed as cost-effective alternatives. Among them is antenna selection (AS) \cite{Win2004_MM,Sanayei2004_CM,Gao2018_CM}, a technique to activate an antenna subset for communications. It is proved that AS is capable of reducing the number of RF chains while simultaneously preserving the DoFs offered by the large-scale antenna array \cite{Win2004_MM}.

In AS-aided MIMO systems, an intriguing question to ask is ``\emph{how the channel capacity scales as the BS antenna number grows large?}'', namely the capacity scaling law (CSL) or the capacity scaling rate. The CSL plays a critical role in defining the capacity limits of AS in MIMO channels, which, thus, provides important system design insights. Because of its importance and extensive applications, the CSL has been one of the research focuses in AS for recent years \cite{Bai2009_TIT,Molisch_TWC_2005,Gao2018_TSP,Li2014_TCOM,Sanayei2007_TIT,Asaad2018_TWC,Park2008_TSP}.
\subsection{Related Works}
In a single-user (SU) MIMO downlink channel where a single transmit antenna with the strongest channel gain is selected, the capacity is shown to scale with the BS antenna number at a double logarithmic rate \cite{Bai2009_TIT}. This work is extended to another downlink case with a single-antenna receiver served by a multi-antenna BS, where a number of BS antennas with strongest channel coefficients are selected for communications \cite{Li2014_TCOM}. Essentially, these two works characterize the CSL achieved by AS in large-scale multiple-input single-input (MISO) systems. As a further advance, with the aid of a well-known capacity bound \cite{Molisch_TWC_2005}, the CSL is unveiled for SU-MIMO uplink systems using the sub-array switching (SAS) architecture, where multiple receive antennas are selected via an exhaustive search (ES). The so-called SAS architecture means that each RF chain is connected to one disjoint sub-array of the BS, which is in contrast to the full-array switching (FAS) architecture where each RF chain is connected to all the BS antennas. Moreover, the CSL achieved by the greedy search-based method \cite{Sanayei2007_TIT} and the channel gains-based method \cite{Asaad2018_TWC} were also studied for SU-MIMO channels. It is noteworthy that the aforementioned works did not consider the CSL in multiuser systems. In an effort to fill this gap, the CSL was discussed for multiuser MIMO multicast channels where multiple antennas are selected via the ES-based method \cite{Park2008_TSP}.
\subsection{Motivations and Main Contributions}
Although the above works have laid a solid foundation for understanding the CSL in massive MIMO systems, they can be improved from three aspects. Firstly, multiuser unicast transmission is one of the most typical applications of massive MIMO, while the CSL therein has not yet been fully understood when multiple antennas are selected. Secondly, some existing CSLs were derived for either specific switching architectures \cite{Gao2018_TSP} or specific selection algorithms \cite{Sanayei2007_TIT,Asaad2018_TWC}, which makes the analytical results lack sufficient generality. Last but not least, most existing works assumed the channels suffer from Rayleigh fading \cite{Bai2009_TIT,Molisch_TWC_2005,Gao2018_TSP,Li2014_TCOM,Sanayei2007_TIT,Asaad2018_TWC,Park2008_TSP}. Although Rayleigh is the most popular fading model, a more accurate fading model for practical radio environment is the Nakagami-$m$ fading that encompasses Rayleigh as a special case \cite{Glodsmith2005}. Nevertheless, the CSL under this fading model has received little attention.

In an effort to overcome these limitations, this article investigates the CSL in multiuser massive MIMO uplink channels when multiple BS antennas are selected by the ES-based method. Under the Nakagami-$m$ fading model, we derive an analytically tractable bound to characterize the sum-rate capacity. By setting the BS antenna number to infinity, we obtain a deterministic equivalent of the sum-rate capacity bound. On this basis, we show that the sum-rate capacity of the considered system grows double logarithmically with the number of BS antennas. It is worth mentioning that the derived CSL is based on the capacity-optimal FAS architecture and ES-based method \cite{Gao2018_TSP}, which, hence, serves as an upper bound for the CSLs achieved by other switching architectures and selection methods.

\subsection{Notation}
Scalars, vectors, and matrices are denoted by non-bold, bold lower-case, and bold upper-case letters, respectively; $\mathbbmss{C}$ stands for the complex plane. The Hermitian of matrix $\mathbf{A}$ is indicated with ${\mathbf{A}}^{\mathsf H}$ and ${\mathbf{I}}_N$ is the $N{\times}N$ identity matrix; $\det(\cdot)$, ${\mathbbmss{E}}\left\{\cdot\right\}$, and ${\mathbbmss{V}}\left\{\cdot\right\}$ denote the determinant, expectation, and variance operator, respectively; ${\rm{erf}}\left(x\right)=\frac{2}{\sqrt{\pi}}\int_{0}^{x}\exp\left(-t^2\right){\rm d}t$ denotes the Gauss error function. In addition, the real Gaussian distribution having mean $\eta$ and variance $\sigma^2$ is represented by ${\mathcal{N}}(\eta,\sigma^2)$. Finally, ${\mathbf x}\sim{\mathcal{CN}}\left({\mathbf 0},{\mathbf X}\right)$ denotes a circularly symmetric complex Gaussian vector having zero mean and covariance matrix $\mathbf X$.

\section{System Model}
\label{section2}
\subsection{Uplink Massive MIMO Channel}
The considered massive MIMO uplink communication system consists of $K$ users and one BS. We assume that each user $k\in{\mathcal{K}}=\left\{1,\ldots,K\right\}$ is equipped with a single transmit antenna to convey signals while the BS has $N\gg K$ antennas for receiving. The received vector at the BS is given by
\begin{align}\label{System_Model}
{{\mathbf{y}}}=\sqrt{p_u}{{\mathbf{G}}}\mathbf{x}+{{\mathbf{n}}}=\sqrt{p_u}{\mathbf{H}}{\mathbf{D}}^{{1}/{2}}\mathbf{x}+{{\mathbf{n}}},
\end{align}
where ${\mathbf{x}}\in{\mathbbmss{C}}^{K\times1}$ is the vector of symbols transmitted by all users and the input covariance matrix is given by ${\mathbf R}_{x}={\mathbbmss E}\left\{{\mathbf{x}}{\mathbf{x}}^{\mathsf{H}}\right\}={\mathbf I}_{K}$; $p_u$ is the average transmit power of each user terminal; ${\mathbf{H}}=\left[{h}_{n,k}\right]\in{{\mathbbmss{C}}^{{N}{\times}{K}}}$ and $h_{n,k}$ models the fast fading from user $k$ to the $n$th antenna of the BS; ${\mathbf{D}}={\mathsf{diag}}\left\{{\beta_1},\ldots,{\beta_{K}}\right\}\in{\mathbbmss{C}}^{K\times K}$ and $\beta_k$ models both geometric attenuation and shadowing from user $k$ to the BS; ${\mathbf G}={\mathbf{H}}{\mathbf{D}}^{{1}/{2}}=\left[g_{n,k}\right]\in{\mathbbmss{C}}^{N\times K}$ is the channel matrix; ${{\mathbf{n}}}\sim{\mathcal{CN}}{({\mathbf{0}},\sigma^2{{\mathbf{I}}_{N}})}$ is the additive white Gaussian noise (AWGN) with $\sigma^2$ being the noise power.

The diagonal elements of $\mathbf D$, i.e., $\{\beta_k\}_{k=1}^{K}$, are assumed to be constant across the antenna array, whereas the elements of ${\mathbf H}$ are assumed to be independent but not identically distributed (i.n.i.d.) with Nakagami-$m$ distributed magnitude and uniformly distributed phase on $[0,2\pi)$. Particularly, the probability density function (PDF) of $\left|h_{n,k}\right|^2$ is given by
\begin{equation}\label{Basic_PDF}
f_{n,k}\left(x\right)=\frac{1}{\Gamma\left(m_k\right)}m_k^{m_j}x^{m_k-1}\exp\left(-m_kx\right), x\geq0,
\end{equation}
where $m_k\geq0.5$ \cite{Glodsmith2005} and $\Gamma\left(z\right)=\int_{0}^{\infty}t^{z-1}\exp\left(-t\right){\rm d}t$ is the complete gamma function. To evaluate the theoretical system performance upper bound, we assume that the CSI is perfectly known at the BS. By denoting $\rho_u=\frac{\rho_u}{\sigma^2}$, the sum-rate capacity of the considered system is given by
\begin{align}
{\mathcal{R}}=\log_2\det\left({{\mathbf{I}}_{N}}+\rho_u{\mathbf{G}}{{\mathbf{G}}^{\mathsf{H}}}\right).
\end{align}

\subsection{Receive Antenna Selection}
Since the BS has full channel information of $\mathbf{G}$, it can exploit receive antenna selection to reduce the hardware implementation complexity. To
guarantee spatial multiplexing gain for the $K$ users, we assume that $L\geq K$ antennas are selected and each antenna is fed by an independent RF chain. Moreover, we assume that an FAS-based architecture is adopted by the BS, where each RF chain is connected to all the BS antennas via a switching network \cite{Gao2018_TSP}. Let ${{\overline{\mathbf{G}}}}\in{{\mathbbmss{C}}^{{L}{\times}{K}}}$ denote the resulting sub-matrix of $\mathbf G$ after receive antenna selection. Without loss of generality, we resort to the sum-rate maximization as the criterion of antenna selection. As a result, the optimal channel sub-matrix satisfies
\begin{align}
{\overline{\mathbf G}}_{\rm{opt}}=\arg\max\nolimits_{{\overline{\mathbf{G}}}{\in}{\mathcal{S}}}\log_2\det\left({{\mathbf{I}}_{L}}+{\rho_u}{\overline{\mathbf{G}}}{{\overline{\mathbf{G}}}}^{\mathsf H}\right),
\label{EQU4}
\end{align}
where $\mathcal{S}$ denotes the full set of the candidate channel sub-matrices with $\left|\mathcal{S}\right|=\binom{N}{L}$. For simplicity, we define $\mathcal{R}_{\rm o}^{(K)}\triangleq\log_2\det\left({{\mathbf{I}}_{L}}+{\rho_u}{\overline{\mathbf{G}}}_{\rm{opt}}{\overline{\mathbf{G}}}{_{\rm{opt}}^{\mathsf H}}\right)$. It is worth noting that problem \eqref{EQU4} is NP-hard and ${\overline{\mathbf G}}_{\rm{opt}}$ can be obtained via the ES-based method with exponential searching complexity ${\mathcal{O}}\left(N^L\right)$.

\section{Sum-Rate Capacity Upper Bound}
While the previous section has established the fundamental model of the massive MIMO systems with receive antenna selection, in the following sections we discuss more properties of the sum-rate capacity therein in order to unveil more system insights. It is worth noting that the calculation of $\mathcal{R}_{\rm o}^{(K)}$ requires an ES, which makes the subsequent analyses of $\mathcal{R}_{\rm o}^{(K)}$ intractable. As a compromise, our efforts will focus on an upper bound of $\mathcal{R}_{\rm o}^{(K)}$. This sum-rate capacity bound is obtained by firstly relaxing the system model described in \eqref{System_Model} to the aggregation of $K$ independent single-input multiple-output sub-channels and then selecting the best $L$ out of $N$ antennas for maximal-ratio combining in each sub-channel. Afterwards, $\mathcal{R}_{\rm o}^{(K)}$ will be upper bounded by the summation of the capacity of the $K$ sub-channels \cite{Molisch_TWC_2005}. Particularly, let $\left\{|\bar{h}_{n,k}|\right\}_{n=1}^{N}$ denote the ordered set of $\left\{|{h}_{n,k}|\right\}_{n=1}^{N}$, i.e., $|\bar{h}_{1,k}|\geq |\bar{h}_{2,k}|\geq\cdots\geq |\bar{h}_{N,k}|$. Then, the sum-rate capacity upper bound can be expressed as
\begin{align}\label{equation9}
{\mathcal{R}}_{\rm{u}}\triangleq\sum\nolimits_{k=1}^{K}\log_2\left(1+{\rho_u\beta_k}{\sum\nolimits_{l=1}^{L}|\bar{h}_{l,k}|^2}\right)
\overset{\star}{\geq}\mathcal{R}_{\rm o}^{(K)},
\end{align}
where the equality in $\star$ holds for $K=1$.

Having defining the upper bound of $\mathcal{R}_{\rm o}^{(K)}$, we now move to characterizing its statistical properties by analyzing its mean for the sake of gleaning further insights. As stated before, the PDF of $\left|h_{n,k}\right|^2$ is given by $f_{n,k}\left(x\right)=\frac{1}{\Gamma\left(m_k\right)}m_k^{m_j}x^{m_k-1}\exp\left(-m_kx\right)$, $x\geq0$. Thus, the joint PDF of the ordered variables $\left\{x_{n,k}=|\bar{h}_{n,k}|^2\right\}_{n=1}^{N}$ is given by
\begin{align}
f_{\bar{h}_k}\left(x_{1,k},\cdots,x_{N,k}\right)=N!\prod\nolimits_{n=1}^{N}
f_{n,k}\left(x_{n,k}\right).
\end{align}
Accordingly, the characteristic function of ${\mathcal{R}}_k\triangleq\log_2\left(1+{\rho_u\beta_k}\sum_{l=1}^{L}{x_{l,k}}\right)$ can be expressed as
\begin{equation}\label{equation8}
\begin{split}
\Phi_{k}\left(\omega\right)=&\int_{0}^{\infty}{\rm d}x_{1,k}\int_{0}^{x_{1,k}}{\rm d}x_{2,k}\cdots\int_{0}^{x_{N-1,k}}{\rm d}x_{N,k}\\
&\times\exp\left({\rm j}\omega{\mathcal{R}}_k\right)N!\prod\nolimits_{n=1}^{N}
f_{n,k}\left(x_{n,k}\right).
\end{split}
\end{equation}
With the characteristic function at hand, the mean of ${\mathcal{R}}_{\rm{u}}$ can be numerically calculated, which yields 
\begin{align}\label{equation_mean}
{\mathbbmss E}\left\{{\mathcal{R}}_{\rm{u}}\right\}=\sum\nolimits_{k=1}^{K}\left.\frac{1}{\rm j}\frac{{\rm d}}{{\rm d}\omega}\Phi_{k}\left(\omega\right)\right|_{\omega=0}. 
\end{align}
Nevertheless, we notice that although \eqref{equation_mean} can be used to calculate ${\mathbbmss E}\left\{{\mathcal{R}}_{\rm{u}}\right\}$, the $(NK)$-fold integrations therein involve a huge computation burden, especially in massive MIMO settings where $N$ is a large value. Furthermore, due to the mathematical intractability of \eqref{equation_mean}, it is challenging to use this characteristic function to unveil more system insights. To handle this difficulty, we intend to derive more concise expressions to approximate ${\mathbbmss E}\left\{{\mathcal{R}}_{\rm{u}}\right\}$, based on which the sum-rate capacity scaling law is explored.

Let us introduce some key preliminary results that will be useful in constructing the approximation of ${\mathbbmss E}\left\{{\mathcal{R}}_{\rm{u}}\right\}$.
\vspace{-5pt}
\begin{lemma}
\label{lemma0}
Let $x_1,x_2,\cdots,x_{N}$ be independent and identically distributed, each with the cumulative distribution function (CDF) $F\left(x\right)$ that satisfies: 1) $\lim_{x\rightarrow+\infty}[F\left(x\right)-F\left(-x\right)]=1$, and 2) $\forall\alpha>0$, $\sup\left\{x:F\left(x\right)\leq\alpha\right\}=\inf\left\{x:F\left(x\right)\geq\alpha\right\}$. Here, $\sup{\mathcal A}$ and $\inf{\mathcal A}$ return the supremum and infimum of set $\mathcal A$, respectively. Let $x_{\left(1\right)}\geq x_{\left(2\right)}\geq\cdots\geq x_{\left(N\right)}$ denote the order statistics of the sample $\left\{x_l\right\}_{l=1,2,\cdots,N}$. Then as $N\rightarrow+\infty$ and for fixed L ($L<N$), $Y_{N}^{(L)}=\sum_{l=1}^{L}x_{\left(l\right)}$ converges in distribution to $Y\sim{\mathcal N}\left(\mu_y,\sigma_y^2\right)$, where
\begin{align}
&\mu_y=N_{\text r}\int_{\upsilon}^{\infty}x{\rm d}F\left(x\right),\\
&\sigma_y^2=L\left(\sigma^2+\left(\upsilon-\frac{\mu_{{y}}}{L}\right)^2\left(1-\frac{L}{N}\right)\right),
\end{align}
and where
\begin{align}\label{Stigler_Con_2}
F\left(\upsilon\right)=1-\frac{L}{N},\quad\sigma^2=\frac{N}{L}\int_\upsilon^{\infty}x^2{\rm d}F\left(x\right)-\frac{\mu{_{y}^2}}{L^2}.
\end{align}
\end{lemma}
\begin{IEEEproof}
Please refer to \cite[pp. 473--475]{Stigler1973}.
\end{IEEEproof}
We now intend to leverage the asymptotic properties stated in Lemma \ref{lemma0} to investigate the statistical features of ${\mathcal{R}}_{\rm{u}}$ in the large limit of the array size, $N$. Particularly, based on Lemma \ref{lemma0}, $\sum_{l=1}^{L}{x_{l,k}}$ converges in distribution to a Gaussian
random variable $X_{k}\sim\mathcal{N}\left(\mu_{k},\sigma_{k}^2\right)$ as $N$ approaches infinity, where
\begin{align}
&{\sigma_{k}^2}=L\left({\bar{\sigma}}_k^2+\left({\upsilon_k}-\frac{\mu_{k}}{L}\right)^2\left(1-\frac{L}{N}\right)\right),\label{Basic_Asymptotic_2}\\
&{\bar{\sigma}}_k^2=\frac{N}{L}\int_{\upsilon_k}^{\infty}x^2{f_{n,k}\left(x\right)}{\rm{d}}x-\frac{\mu{_{k}^2}}{L^2}\\
&\quad=\frac{N}{Lm_k^2\Gamma\left(m_k\right)}\Upsilon\left(m_k+2,m_k{\upsilon_k}\right)-\frac{\mu{_{k}^2}}{L^2},\label{Basic_Asymptotic_3}
\end{align}
\begin{align}
&\mu_{k}=N\int_{\upsilon_k}^{\infty}x{f_{n,k}\left(x\right)}{\rm{d}}x
=\frac{N\Upsilon\left(m_k+1,m_k{\upsilon_k}\right)}{m_k\Gamma\left(m_k\right)},\label{Basic_Asymptotic_1}\\
&\int_{{\upsilon_k}}^{\infty}f_{n,k}\left(x\right){\rm{d}}x=\frac{1}{\Gamma\left(m_k\right)}\Upsilon\left(m_k,m_k\upsilon_k\right)=\frac{L}{N},\label{Basic_Asymptotic_4}
\end{align}
and where $\Upsilon\left(s,x\right)=\int_{x}^{+\infty}t^{s-1}\exp\left(-t\right){\rm d}t$ is the upper incomplete gamma function. By defining ${\mathcal{R}}_{\rm{a}}\triangleq\sum_{k=1}^{K}\log_2\left(1+{\rho_u\beta_k}X_k\right)$, we can obtain
\begin{equation}
\lim\nolimits_{N\rightarrow\infty}{\mathbbmss E}\left\{{\mathcal{R}}_{\rm{u}}\right\}={\mathbbmss E}\left\{{\mathcal{R}}_{\rm{a}}\right\}.
\end{equation}

Based on Lemma \ref{lemma0}, ${\mathbbmss E}\left\{{\mathcal{R}}_{\rm{u}}\right\}={\mathbbmss E}\left\{{\mathcal{R}}_{\rm{a}}\right\}$ holds in the large limit of $N$. However, as will be shown in Section \ref{section5}, ${\mathbbmss E}\left\{{\mathcal{R}}_{\rm{a}}\right\}$ can provide an accurate approximation of ${\mathbbmss E}\left\{{\mathcal{R}}_{\rm{u}}\right\}$ even for finite system dimensions, such as $N=128$. As a result, the sum-rate capacity upper bound can be approximated as ${\mathbbmss E}\left\{{\mathcal{R}}_{\rm{u}}\right\}\approx{\mathbbmss E}\left\{{\mathcal{R}}_{\rm{a}}\right\}$ \cite{Ouyang2019_ICC,Ouyang2020_TCOM}, which suggests that
\begin{align}
&{\mathbbmss E}\left\{{\mathcal{R}}_{\rm{u}}\right\}\approx
\sum\nolimits_{k=1}^{K}\int_{-\infty}^\infty{\log_2\left(1+{\rho_u\beta_k}x\right)f_{X_k}\left(x\right){\rm{d}}x}\label{Upper_Bound2_Mean},
\end{align}
where $f_{X_k}\left(x\right)=\frac{1}{\sqrt{2\pi\sigma_{k}^2}}
\exp\left({-\frac{\left(x-\mu_{k}\right)^2}{2\sigma_{k}^2}}\right)$.

It is worth mentioning that the random variable $\sum_{l=1}^{L}{x_{l,k}}=\sum_{l=1}^{L}|\bar{h}_{l,k}|^2$ takes positive values for sure. Yet, its approximation, $X_{k}\sim\mathcal{N}\left(\mu_{k},\sigma_{k}^2\right)$, can take negative values for a small probability. This is due to the fact that $X_k$ is only an approximation of $\sum_{l=1}^{L}{x_{l,k}}$ when $N<\infty$. As will be detailed in Appendix, $\lim_{N\rightarrow\infty}\mu_{k}=\infty$ and $\lim_{N\rightarrow\infty}\frac{\sigma_{k}^2}{\mu_k^2}=0$, which indicates that
\begin{align}
\lim_{N\rightarrow\infty}\Pr\left(X_{k}<0\right)&=\lim_{N\rightarrow\infty}\left[1+{\rm{erf}}\left(-{\mu_{k}^2}/{\sigma_{k}^2}\right)\right]/2\\
&=\left[1+{\rm{erf}}\left(-\infty\right)\right]/2=0.
\end{align}
The above arguments imply that the deviation of the approximated distribution $\mathcal{N}\left(\mu_{k},\sigma_{k}^2\right)$ from the exact distribution of $\sum_{l=1}^{L}{x_{l,k}}$ vanishes as $N$ grows infinitely large. Considering this fact, we can also approximate ${\mathbbmss E}\left\{{\mathcal{R}}_{\rm{u}}\right\}$ as
\begin{align}
{\mathbbmss E}\left\{{\mathcal{R}}_{\rm{u}}\right\}\approx
\sum\nolimits_{k=1}^{K}\int_{-\infty}^\infty{\log_2\left(1+{\rho_u\beta_k}|x|\right)f_{X_k}\left(x\right){\rm{d}}x}.\label{Upper_Bound2_Mean1}
\end{align}
We notice that both \eqref{Upper_Bound2_Mean} and \eqref{Upper_Bound2_Mean1} only involve $K$-fold integrations, which, thus, are much more computationally tractable than the characteristic function presented in \eqref{equation8}.

\begin{figure*}[!t]
    \centering
    \subfigbottomskip=0pt
	\subfigcapskip=-5pt
\setlength{\abovecaptionskip}{0pt}
   \subfigure[Sum-rate capacity versus the transmit power $p_u$.]
    {
        \includegraphics[height=0.24\textwidth]{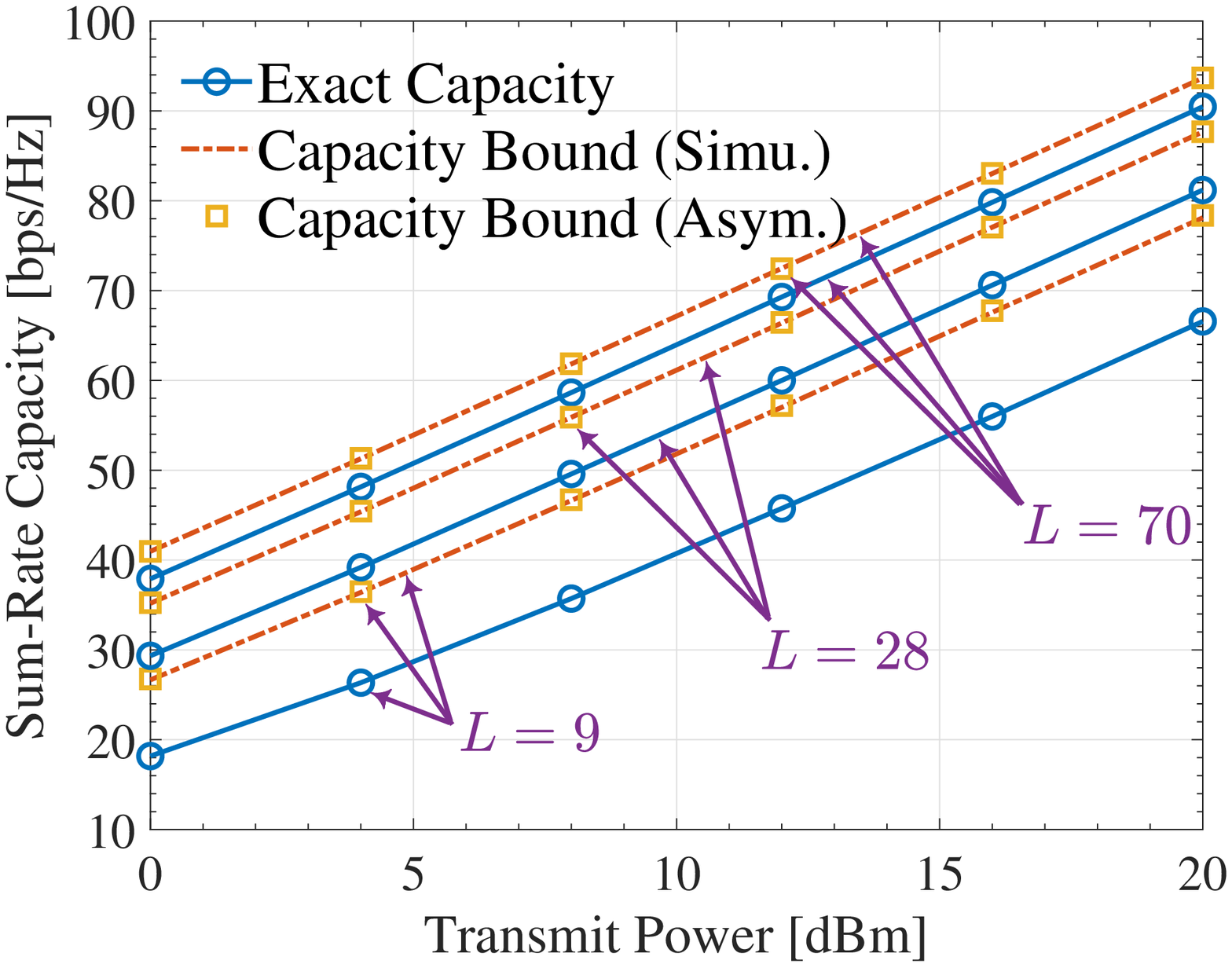}
	   \label{fig1a}	
    }
    \subfigure[The mean of the sum-rate capacity bound.]
    {
        \includegraphics[height=0.24\textwidth]{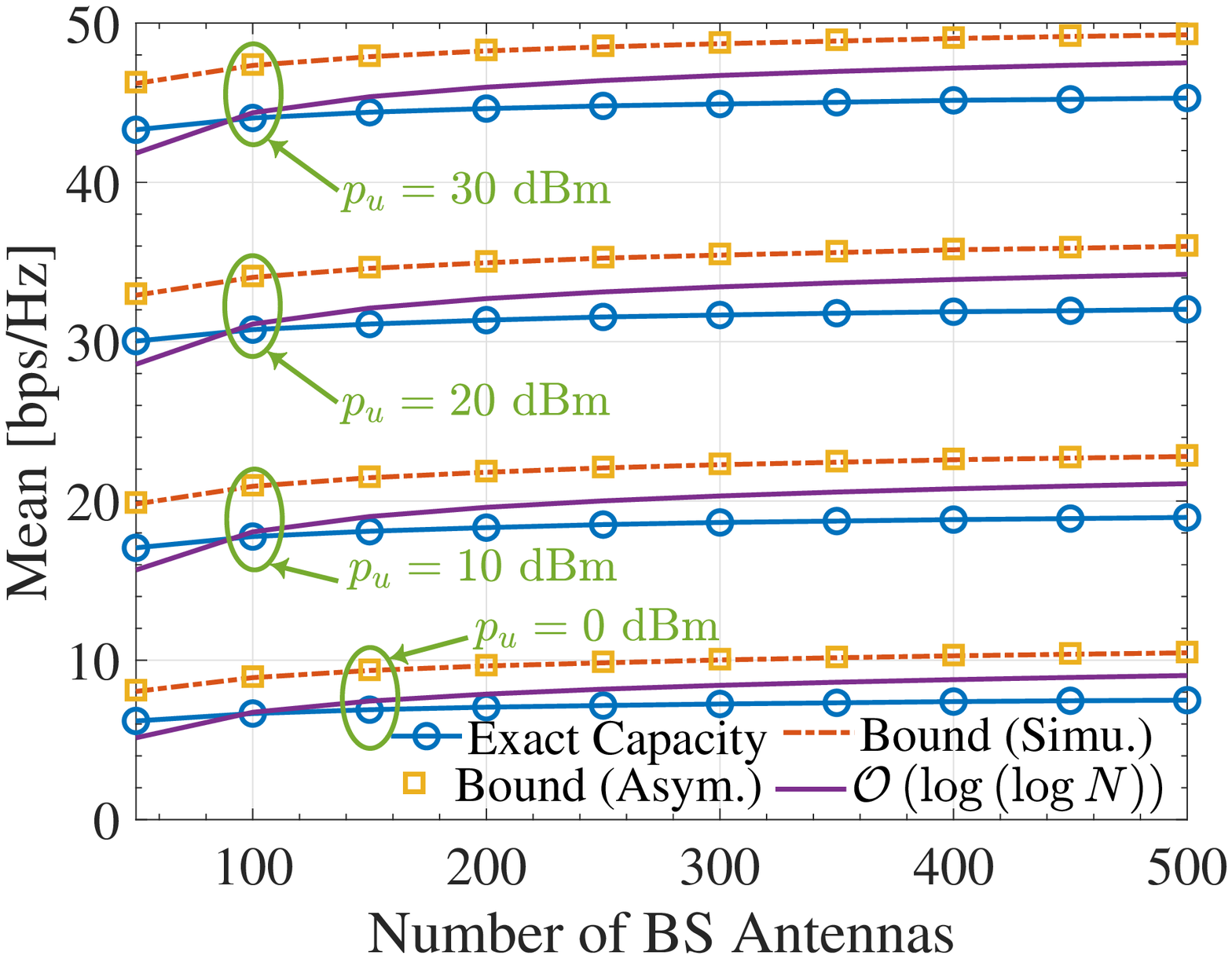}
	   \label{fig1b}	
    }
    \subfigure[The SCV of the capacity bound.]
    {
        \includegraphics[height=0.24\textwidth]{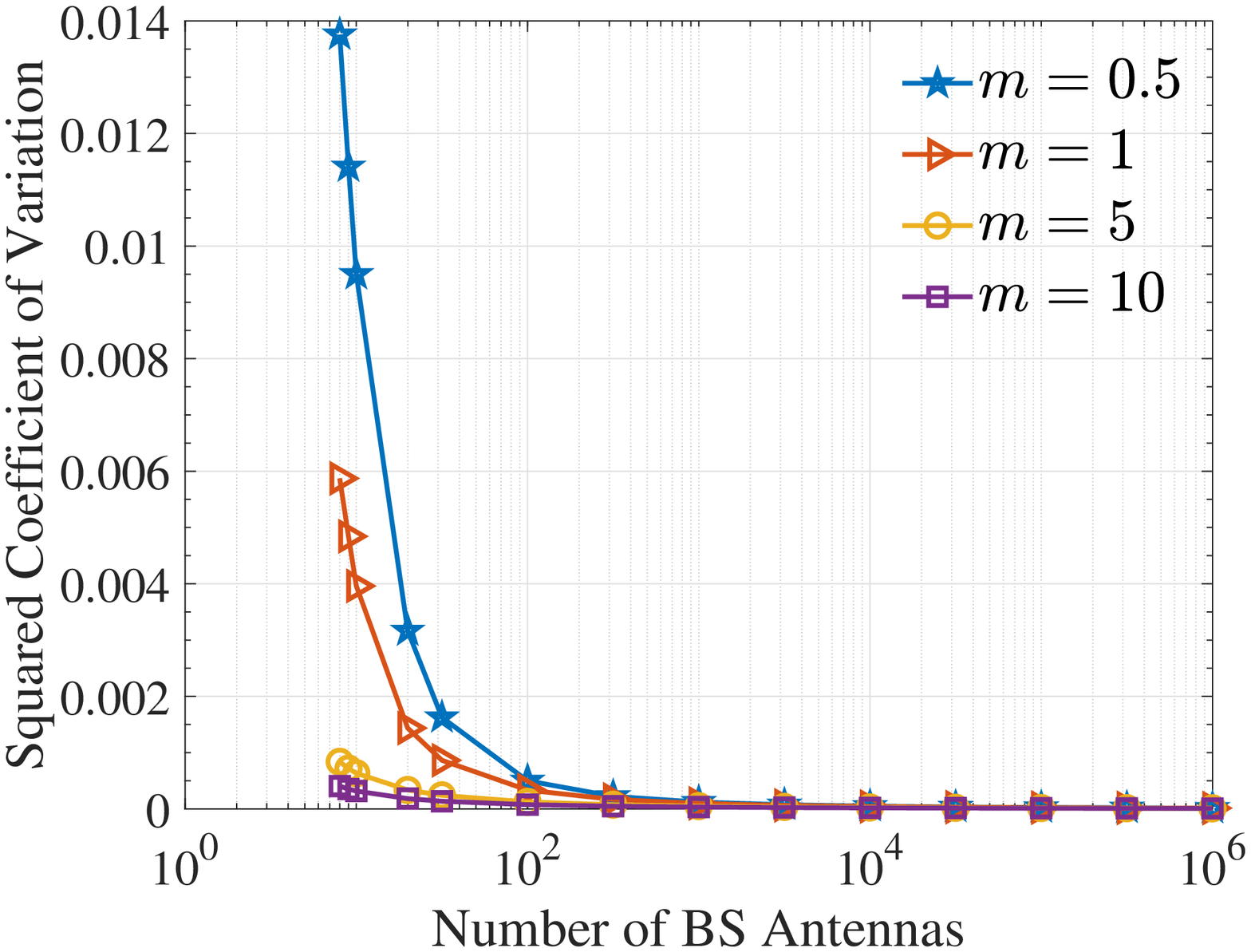}
	   \label{fig1c}	
    }
   \caption{Sum-rate capacity and its upper bound of the considered system. The simulation parameters are given as follows: (a) $N=128$ and $K=8$; (b) $K=4$ and $L=8$; (c) $K=4$, $L=8$, and $p_u=10$ dBm.}
    \label{figure1}
    \vspace{-20pt}
\end{figure*}

\section{Sum-Rate Capacity Scaling Law}
The fact of $\lim\nolimits_{N\rightarrow\infty}{\mathbbmss E}\left\{{\mathcal{R}}_{\rm{u}}\right\}={\mathbbmss E}\left\{{\mathcal{R}}_{\rm{a}}\right\}$ suggests that ${\mathbbmss E}\left\{{\mathcal{R}}_{\rm{a}}\right\}$ presents the same scaling law as ${\mathbbmss E}\left\{{\mathcal{R}}_{\rm{u}}\right\}$ with the increment of $N$. Accordingly, it makes sense to rely on ${\mathbbmss E}\left\{{\mathcal{R}}_{\rm{a}}\right\}$ to explore the scaling law of ${\mathbbmss E}\left\{{\mathcal{R}}_{\rm{u}}\right\}$. By setting $N$ as infinity, the following theorem can be found.
\vspace{-5pt}
\begin{theorem}
\label{theorem2}
For fixed $L$ ($L<N$), $\lim\limits_{N\rightarrow\infty}\frac{{\mathbbmss E}\left\{{\mathcal{R}}_{\rm{u}}\right\}}{\sum_{k=1}^{K}\log_2\left(1+{L}\frac{\rho_u\beta_k}{m_k}{\log{{N}}}\right)}=1$ and $\lim\limits_{N\rightarrow\infty}\frac{{\mathbbmss V}\left\{{\mathcal{R}}_{\rm{u}}\right\}}{({\mathbbmss E}\left\{{\mathcal{R}}_{\rm{u}}\right\})^2}=0$.
\end{theorem}
\vspace{-5pt}
\begin{IEEEproof}
The proof is detailed in Appendix.
\end{IEEEproof}
Note that in statistics, $\frac{{\mathbbmss V}\left\{{\mathcal{R}}_{\rm{u}}\right\}}{({\mathbbmss E}\left\{{\mathcal{R}}_{\rm{u}}\right\})^2}$ is also termed as the of the squared coefficient of variation (SCV) of the random variable ${\mathcal{R}}_{\rm{u}}$. Based on Theorem \ref{theorem2}, we can write ${\mathcal{R}}_{\rm{u}}$ as 
\begin{align}
{\mathcal{R}}_{\rm{u}}={\mathcal{X}}\sum\nolimits_{k=1}^{K}\log_2\left(1+{\rho_u\beta_k}\frac{L}{m_k}{\log{{N}}}\right) 
\end{align}
with the random variable $\mathcal{X}$ satisfying $\lim_{N\rightarrow\infty}\mathbbmss{E}\{{\mathcal{X}}\}=1$ and $\lim_{N\rightarrow\infty}\mathbbmss{V}\{{\mathcal{X}}\}=0$. Therefore, it can be easily shown that as $N\rightarrow\infty$,
\begin{align}\label{Deterministic_Equivalent}
{\mathcal{R}}_{\rm{u}}-\sum\nolimits_{k=1}^{K}\log_2\left(1+{\rho_u\beta_k}\frac{L}{m_k}{\log{{N}}}\right)
\overset{a.s.}{\longrightarrow}0,
\end{align}
where -a.s. denotes almost sure convergence. Note that $\sum_{k=1}^{K}\log_2\left(1+{\rho_u\beta_k}\frac{L}{m_k}{\log{{N}}}\right)$ serves as a deterministic equivalent for ${\mathcal{R}}_{\rm{u}}$.
\vspace{-5pt}
\begin{corollary}\label{Capacity_Scaling_Law}
The sum-rate capacity, $\mathcal{R}_{\rm o}^{(K)}$, scales with $N$ at a double logarithmic rate.
\end{corollary}
\vspace{-5pt}
\begin{IEEEproof}
The results in \eqref{Deterministic_Equivalent} indicate that $\mathcal{R}_{\rm u}=\mathcal{O}\left(\log(\log{N})\right)$ as $N\rightarrow\infty$. Let $\mathcal{R}_1$ denote the capacity bound for $K=1$. Thus, we have $\mathcal{R}_{1}=\mathcal{O}\left(\log(\log{N})\right)$ and $\mathcal{R}_{\rm u}\geq \mathcal{R}_{\rm o}^{(K)}\geq\mathcal{R}_{\rm o}^{(1)}= {\mathcal{R}}_1$. Using the Sandwich Theorem, we get $\mathcal{R}_{\rm o}^{(K)}=\mathcal{O}\left(\log(\log{N})\right)$ and the proof is completed.
\end{IEEEproof}
\vspace{-5pt}
\begin{remark}
The results in \eqref{Deterministic_Equivalent} indicate that ${\mathcal{R}}_{\rm{u}}$ converges to $\sum_{k=1}^{K}\log_2\left(1+{\rho_u\beta_k}\frac{L}{m_k}{\log{{N}}}\right)$ with probability 1 in the large system limit, which can be treated as a consequence of the channel hardening \cite{Hochwald2004}.
\end{remark}
\vspace{-5pt}
\vspace{-5pt}
\begin{remark}
The results in \eqref{Deterministic_Equivalent} also indicate that a larger value of the sum-rate capacity bound can be attained in a richer scattering environment with a smaller Nakagami parameter $m$ \cite{Glodsmith2005}. This is because that more gains are possible due to antenna selection for smaller values of $m$. More specifically, due to the larger channel variations involved by richer scattering, the system sum-rate can be more improved by properly choosing a candidate subsect of antennas, and thus a smaller value of $m$ yields a larger selection diversity gain. When $m$ tends to infinity, the channel becomes the AWGN channel and no selection diversity gain can be exploited because there is no channel variation.
\end{remark}
\vspace{-5pt}
\vspace{-5pt}
\begin{remark}
Note that ${\mathcal{R}}_{\rm{u}}$ is achieved by the FAS-based architecture and the ES-based selection method, both being capacity-optimal. Hence, it can be concluded that the sum-rate achieved by other switching architectures and selection methods will grow no faster than double logarithmically with the BS antenna number.
\end{remark}
\vspace{-5pt}

\section{Numerical Results}\label{section5}
In this section, numerical results are provided to verify the correctness of the above theoretical analyses. During the simulations, the user terminals are assumed to be distributed uniformly over a hexagonal cell with radius of 1000 m. The free-space path loss model is $-10\log_{10}\beta_k=92.5+20\log_{10}\left[f_0\left(\text{GHz}\right)\right]+20\log_{10}\left[d_k\left(\text{km}\right)\right]$, where $f_0=4$ GHz is the carrier frequency and $d_k$ is the distance between the BS and user $k$. Besides, we set $\sigma^2=-100$ dBm. All the simulation results are obtained via $5\times10^5$ Monte-Carlo trials.

To verify the correctness of \eqref{Upper_Bound2_Mean}, {\figurename} {\ref{fig1a}} plots the exact sum-rate capacity, the simulated capacity bound, and the asymptotically approximated capacity bound (calculated by \eqref{Upper_Bound2_Mean}) in terms of the transmit power for selected numbers of $L$. It can be observed that the asymptotically approximated results match perfectly with the simulations, which, thus, verifies our previous derivations. Moreover, the capacity bound becomes tighter for a larger value of $L$. As {\figurename} {\ref{fig1a}} shows, the sum-rate capacity upper bound is tight and presents a similar changing trend to the exact capacity. Accordingly, one can conclude that this bound serves as a good alternative for the exact capacity in terms of system sum-rate performance evaluation. Actually, the Gaussian-distribution approximation stated in Lemma \ref{lemma0} is based on the assumption of $N\rightarrow\infty$. Yet, it can be seen from {\figurename} {\ref{fig1a}} that this approximation works well even for a limited value of $N$, i.e., $N=128$, which indicates the robustness of Lemma \ref{lemma0} as well as the tightness of \eqref{Upper_Bound2_Mean}.

{\figurename} {\ref{fig1b}} shows the mean of the capacity bound versus $N$ for a fixed value of $L$. For reference, the curves representing the exact sum-rate capacity and $\sum_{k=1}^{K}\log_2\left(1+{\rho_u\beta_k}\frac{L}{m_k}{\log{{N}}}\right)$ (labeled as `${\mathcal{O}}\left(\log\left(\log N\right)\right)$') are also plotted. As shown, both the capacity and its upper bound grow like ${\mathcal{O}}\left(\log\left(\log N\right)\right)$, which supports our conclusions in Corollary \ref{Capacity_Scaling_Law}. Then, we assume $m_k=m$, $\forall k\in\mathcal{K}$, and plot the SCV of ${\mathcal{R}}_{\rm{u}}$, i.e., $\frac{{\mathbbmss V}\left\{{\mathcal{R}}_{\rm{u}}\right\}}{({\mathbbmss E}\left\{{\mathcal{R}}_{\rm{u}}\right\})^2}$, versus $N$ for selected values of $m$ in {\figurename} {\ref{fig1c}}. As this graph shows, the SCV tends towards zero as $N$ increases and this consists with the conclusion drawn in Theorem \ref{theorem2}. Moreover, as stated before, a smaller Nakagami-$m$ fading parameter corresponds to larger channel variations, which yields a larger value of the SCV as well as a slower channel hardening speed. This is also validated by {\figurename} {\ref{fig1c}}.

\section{Conclusion}
The sum-rate capacity achieved by receive antenna selection in massive MIMO uplink systems have been discussed. For the sake of mathematical tractability, a low-complexity sum-rate capacity upper bound has been proposed. A deterministic equivalent of this bound has been found to characterize its asymptotic behavior in the large limit of the BS antenna number. On this basis, the sum-rate capacity is shown to sacle with the BS antenna number at a double logarithmic rate.

The results of this study can be used to investigate antenna selection-aided massive MIMO systems in various respects, which will be addressed in the extended version of this work.

\begin{appendix}
It follows from \eqref{Basic_Asymptotic_4} that $\lim_{N\rightarrow\infty}\int_{{\upsilon_k}}^{\infty}f_{n,k}\left(x\right){\rm{d}}x=\lim_{N\rightarrow\infty}\frac{L}{N}=0$ and thus $\lim_{N\rightarrow\infty}\upsilon_k=\infty$. Using \cite[eq. (8.11.2)]{Paris2010}, we have
\begin{align}
\lim_{N\rightarrow\infty}\frac{\int_{{\upsilon_k}}^{\infty}f_{n,k}\left(x\right){\rm{d}}x}{L/N}=\lim_{N\rightarrow\infty}
\frac{{\frac{\left(m_k{\upsilon_k}\right)^{m_k-1}}{\Gamma(m_k)\exp\left({m_k{\upsilon_k}}\right)}}}{L/N}=1, \label{Asymptotic_Analysis_Basic}
\end{align}
which yields 
\begin{align}
\lim_{N\rightarrow\infty}\frac{\log{\frac{\left(m_k{\upsilon_k}\right)^{m_k-1}}{\Gamma(m_k)\exp\left({m_k{\upsilon_k}}\right)}}}
{\log{N}-\log{L}}=\lim_{N\rightarrow\infty}\frac{m_k{\upsilon_k}}{\log{N}}=1. 
\end{align}
Similarly, based on \eqref{Basic_Asymptotic_1}, we can get
\begin{align}
 \lim\limits_{N\rightarrow\infty}\frac{\mu_{k}}{L{\upsilon_k}}&
 =\lim\limits_{N\rightarrow\infty}\frac{\frac{Nm_k^{m_k-1}{\upsilon_k}^{m_k}}{\Gamma\left(m_k\right)\exp\left({m_k{\upsilon_k}}\right)}}{{\upsilon_k} L}
 \overset{\diamond}{=}\lim\limits_{N\rightarrow\infty}\frac{N{\upsilon_k}\frac{L}{N}}{{\upsilon_k} L}=1,\nonumber
\end{align}
where the equality in $\diamond$ holds for \eqref{Asymptotic_Analysis_Basic}. By continuously following the above steps, we have $\lim_{N\rightarrow\infty}\frac{\sigma_{k}^2}{L^2{\upsilon_k}^2}=0$. As stated before, $\lim_{N\rightarrow\infty}\frac{\mathbbmss{E}\{\mathcal{R}_k\}}{\log_2\left(1+{\rho_u\beta_k}X_k\right)}=1$, where $X_{k}\sim\mathcal{N}\left(\mu_{k},\sigma_{k}^2\right)$, which together with the fact of $\lim_{N\rightarrow\infty}\frac{\mu_{k}}{L{\upsilon_k}}=1$, suggests that $\mathbbmss{E}\{\mathcal{R}_k\}$ satisfies
\begin{align}\label{equationA23}
\lim_{N\rightarrow\infty}\frac{\mathbbmss{E}\{\mathcal{R}_k\}}{{\mathbbmss E}\{\log_2\left(1+{\rho_u\beta_k}L{\upsilon_k} W_k\right)\}}=1.
\end{align}
Here, $W_k=\frac{X_k}{L{\upsilon_k}}\sim\mathcal{N}\left(\frac{\mu_{k}}{L{\upsilon_k}},\frac{\sigma_{k}^2}{L^2{\upsilon_k^2}}\right)$ with $\lim_{N\rightarrow\infty}{{\mathbbmss{E}}\left\{W_k\right\}}=1$ and $\lim_{N\rightarrow\infty}{{\mathbbmss{V}}\left\{W_k\right\}}=0$. Hence, the random variable $W_k$ converges to 1 with probability one as $N\rightarrow\infty$, which together with \eqref{equationA23} and the fact of $\lim_{N\rightarrow\infty}\frac{{\upsilon_k}}{\frac{1}{m_k}\log{N}}=1$, suggests that $\lim_{N\rightarrow\infty}\frac{\mathbbmss{E}\{\mathcal{R}_k\}}{\log_2\left(1+{\rho_u\beta_k}L\frac{1}{m_k}\log{N}\right)}=1$. Note that for $\forall a,b>0$, it has $\lim_{x\rightarrow\infty}\frac{\log\left(1+ax\right)}{\log\left(1+bx\right)}=1$. Therefore, we can obtain 
\begin{align}\label{DE_Result1}
\lim_{N\rightarrow\infty}\frac{\sum_{k=1}^{K}\mathbbmss{E}\{\mathcal{R}_k\}}{\sum_{k=1}^{K}\log_2\left(1+{\rho_u\beta_k}L\frac{1}{m_k}\log{N}\right)}=1,
\end{align}
which completes the proof of the first part of Theorem \ref{theorem2}. Following a similar approach as that in obtaining \eqref{DE_Result1}, we find that ${\mathcal{P}}_k=\frac{{\mathbbmss V}\left\{{\mathcal{R}}_{k}\right\}}{({\mathbbmss E}\left\{{\mathcal{R}}_{k}\right\})^2}$ satisfies
\begin{align}\label{Equation61}
 \lim\limits_{N\rightarrow\infty}{\mathcal{P}}_k=
 \lim\limits_{N\rightarrow\infty}\frac{{\mathbbmss{E}}\left\{\left[\log_2\left(1+{\rho_u\beta_k}L{\upsilon_k} W_k\right)\right]^2\right\}}{({\mathbbmss E}\left\{{\mathcal{R}}_{k}\right\})^2}-1.\nonumber
 \end{align}
Since $\lim_{N\rightarrow\infty}{\mathbbmss E}\left\{W_k\right\}=1$, $\lim_{N\rightarrow\infty}{\mathbbmss V}\left\{W_k\right\}=0$, and $\lim_{N\rightarrow\infty}\frac{\mathbbmss{E}\{\mathcal{R}_k\}}{{\mathbbmss E}\{\log_2\left(1+{\rho_u\beta_k}L{\upsilon_k}\right)\}}=1$, we get $\lim_{N\rightarrow\infty}{\mathcal{P}}_k=0$. Based on this and the fact of $\lim_{x\rightarrow\infty}\frac{(\log(1+ax))^2}{(\log(1+bx))^2}=1$ ($\forall a,b>0$), we arrive at the following relationship:
\begin{equation}
\begin{split}
 &0\leq\lim\limits_{N\rightarrow\infty}\frac{{\mathbbmss V}\left\{{\mathcal{R}}_{\rm{u}}\right\}}{({\mathbbmss E}\left\{{\mathcal{R}}_{\rm{u}}\right\})^2}=\lim\limits_{N_{\text r}\rightarrow\infty}\frac{\sum_{k=1}^{K}{\mathbbmss V}\left\{{\mathcal{R}}_{k}\right\}}{\left(\sum_{k=1}^{K}{\mathbbmss E}\left\{{\mathcal{R}}_{k}\right\}\right)^2}\\
 &\leq\lim\limits_{N\rightarrow\infty}\frac{\sum_{k=1}^{K}{\mathbbmss V}\left\{{\mathcal{R}}_{k}\right\}}
 {\sum_{k=1}^{K}({\mathbbmss E}\left\{{\mathcal{R}}_{k}\right\})^2}=0.
\end{split}
\end{equation}
Thus, the second part of Theorem \ref{theorem2} is proved.
\end{appendix}

\vspace{12pt}
\end{document}